# Formation of Compressed Flat Electron Beams with High Transverse-Emittance Ratios

J. Zhu[a,b,*], P. Piot[c,d], D. Mihalcea[c], C. R. Prokop[c]

[a]*Accelerator Division, Fermi National Accelerator Laboratory, Batavia, IL 60510, USA*
[b]*Institute of Fluid Physics, CAEP, Mianyang 621900, China*
[c]*Department of Physics, Northern Illinois University DeKalb, IL 60115, USA*
[d]*Accelerator Physics Center, Fermi National Accelerator Laboratory, Batavia, IL 60510, USA*

*Corresponding author*
*Email address: zhujun98@fnal.gov, zhujun981661@gmail.com*

Flat beams – beams with asymmetric transverse emittances – have important applications in novel light-source concepts, advanced-acceleration schemes and could possibly alleviate the need for damping rings in lepton colliders. Over the last decade, a flat-beam-generation technique based on the conversion of an angular-momentum-dominated beam was proposed and experimentally tested. In this paper we explore the production of compressed flat beams. We especially investigate and optimize the flat-beam transformation for beams with substantial fractional energy spread. We use as a simulation example the photoinjector of the Fermilab's Advanced Superconducting Test Accelerator (ASTA). The optimizations of the flat beam generation and compression at ASTA were done via start-to-end numerical simulations for bunch charges of 3.2 nC, 1.0 nC and 20 pC at ~37 MeV. The optimized emittances of flat beams with different bunch charges were found to be 0.25 μm (emittance ratio is ~400), 0.13 μm, 15 nm before compression, and 0.41 μm, 0.20 μm, 16 nm after full compression, respectively with peak currents as high as 5.5 kA for a 3.2-nC flat beam. These parameters are consistent with requirements needed to excite wakefields in asymmetric dielectric-lined waveguides or produce significant photon flux using small-gap micro-undulators.

## I INTRODUCTION

Flat beams with high-emittance ratio directly generated in a photoinjector via linear transformation of angular-momentum-dominated round beams (also referred to as "magnetized beams") have many attracting applications. Owing to its simplicity, this flat-beam photoinjector generation was initially proposed as a potential simple and cost-effective alternative to electron damping rings in an e+/e- linear collider [1]. Since then, other potential uses of flat beam technique have emerged. These applications include (i) tabletop THz source based on beam-grating interaction [the so-called Smith-Purcell free-electron laser (FEL)] [2], (ii) mitigation of collective effects and use in single-pass FEL [3], and (iii) development of beam- [4] or laser-driven [5] accelerating structure using asymmetric dielectric structures. The flat-beam transformation could also be combined with transverse-to-longitudinal emittance exchange techniques, and thereby enable arbitrary repartitioning of the beam emittances within the three degrees of freedom [6]. Experimental demonstration of the flat-beam source have been achieved at the Fermilab A0 Photoinjector Laboratory (A0PI) [7-8], where an emittance ratio of ~100 was achieved for a bunch charge of ~0.5 nC with the smaller emittance of the flat beam of ~0.41 μm.

Compressed low-energy (<100 MeV) flat beams required by, e.g., advanced beam-driven acceleration techniques are challenging to obtain and no careful investigation has been performed to date. Often, electron bunches are accelerated to energies where the space charge (SC) forces are weakened sufficiently (due to their $1/\gamma^2$ dependence, where $\gamma$ is the Lorentz factor), and then longitudinally compressed to achieve the desired peak current [9-10]. Bunch compression at low energy, might be necessary in compact accelerator-based light sources and injector for advanced acceleration concepts. At these energies, high-charge bunches (~nC) might be subject to significant space-charge effects that could lead to substantial emittance dilution. Besides, the bunch self-interaction with its own radiation field via coherent synchrotron radiation (CSR) can significantly impact the beam dynamics [11-16] and results in further phase-space degradations. Simulations of round beams for the ASTA injector indicate that the CSR effect contributes to most of the emittance growth for high-bunch charges, while the space-charge effects are predominant at lower bunch charges [17]. Preliminary simulations of the compression of a flat beam with different incoming transverse emittance ratios indicate that the four-dimensional emittance degradation is mitigated for flat beams with high transverse emittance ratios [17].

In this paper, we report on the optimization of flat beams generation and compression at ~50 MeV or below. Our simulations use the 50-MeV photoinjector of the Advanced Superconducting Test Accelerator (ASTA) [18] currently in the commissioning phase at Fermilab. In Section II we summarize the theory of the photoinjector production of the flat beam and especially discuss chromatic effects and their possible compensation in the flat-beam transformation. Section III outlines the ASTA photoinjector setup. Sections IV and V, respectively present the generation and compression of flat beams. The main results are summarized in Section VI.

## II THEORY OF FLAT BEAM GENERATION

In this paper, the coordinates of an electron in 6D phase space is denoted as $[x, x', y, y', z, \delta]$. Here $z$ denotes the longitudinal distance from the center of the electron bunch, and $\delta=\Delta\gamma/\gamma_0$ denotes the fractional energy spread ($\gamma_0$ is the bunch mean energy and $\Delta\gamma$ the energy offset relative to the mean energy).

The generation of a flat beam in a photoinjector is staged. First a canonical-angular-momentum (CAM) dominated beam is produced and accelerated to relativistic energies. Then a linear transformation henceforth referred to as round-to-flat-beam transformation (RFBT) is applied to remove the angular momentum thereby resulting in a beam with asymmetric transverse emittances.

To produce a CAM-dominated beam, the photocathode is immersed in an axial magnetic field $B_0=B_z(z=0)$, which introduces an average canonical angular momentum:

$$\langle L \rangle = eB_0\sigma_c^2 \quad (1)$$

where $e$ is the charge of an electron, and $\sigma_c$ is the root-mean-square (r.m.s.) transverse size of the laser spot on the cathode. The angular momentum of the beam can be properly removed by a RFBT transformer consisting of three skew quadrupole magnets, and the beam will be transformed into a flat beam with transverse geometric emittances given as [19-20]:

$$\varepsilon_{\pm} = \sqrt{\varepsilon_u^2 + \ell^2} \pm \ell \xrightarrow{\ell \gg \varepsilon_u} \begin{cases} \varepsilon_+ \approx 2\ell \\ \varepsilon_- \approx \dfrac{\varepsilon_u^2}{2\ell} \end{cases} \quad (2)$$

where $\varepsilon_u$ is the uncorrelated geometric emittance, $\ell =\langle L\rangle/2p_z$, and $p_z$ is the longitudinal momentum. The emittances of the flat beam coincide with the eigen-emittances [21] of the CAM-dominated beam, namely:

$$\varepsilon_{\pm} = \varepsilon_{\text{eigen},\pm} \quad (3)$$

where $\varepsilon_{\text{eigen},\pm}$ for the CAM-dominated beam can be found by solving the following equation [22]:

$$\det\left(J\Sigma - i\varepsilon_{\text{eigen},\pm}I\right) = 0 \quad (4)$$

where $I$ and $J$ are respectively the 4×4 unit and unit symplectic matrix, and $\Sigma$ is the 4×4 covariance matrix of the beam. In the following, we will refer to the horizontal emittance of the flat beam $\varepsilon_x=\varepsilon_+$ and the vertical emittance of the flat beam $\varepsilon_y=\varepsilon_-$.

In order to compress the flat beam to produce a high peak current, a large longitudinal-phase-space (LPS) chirp $h = -\langle\delta z\rangle/\langle z^2\rangle$ is produced by off-crest acceleration in the booster cavity downstream of the photocathode. Maximum bunch compression is achieved when $h=-1/R_{56}$. The chromatic aberration during the RFBT transformation will result in considerable emittance growth of the flat beam [23]. The relative flat-beam smaller-emittance growth during the RFBT is derived as (Appendix A):

$$\frac{\Delta\varepsilon_y}{\varepsilon_{y0}} \approx \frac{1}{2}\langle\delta^2\rangle \begin{pmatrix} 2q_{0,1}^2\langle x_{i,1}^2\rangle\langle y_{i,1}^2\rangle + (1+\lambda)q_{0,2}^2\langle x_{i,2}^2\rangle\langle y_{i,2}^2\rangle \\ +2q_{0,3}^2\langle x_{i,3}^2\rangle\langle y_{i,3}^2\rangle - \varepsilon_{\text{eff}}^2 \end{pmatrix} / \varepsilon_{y0}^2$$

(5)

where $q_{0,j}$ is the integral quadrupole strength of the $j^{\text{th}}$ skew-quadrupole magnet, $x_{i,j}$ and $y_{i,j}$ are the coordinates of electrons at the entrance of the $j^{\text{th}}$ skew-quadrupole magnet,

$\varepsilon_{\text{eff}}$ is the geometric emittance of the magnetized beam upstream of the RFBT, $\varepsilon_{y0}$ is the nominal vertical geometric emittance of the flat beam in absence of chromatic aberration and $0\leq\lambda\leq1$. Equation (5) is very similar to what has been derived for the normal quadrupole magnet [24]. An important difference is that the emittance growth during the RFBT is determined by both the horizontal and vertical beam sizes.

To convert a magnetized beam into a flat beam, the following equation should be satisfied [20]:

$$S = A_+^{-1} A_- \quad (6)$$

where

$$S = \begin{bmatrix} \alpha & \beta \\ -\dfrac{1+\alpha^2}{\beta} & -\alpha \end{bmatrix} \quad (7)$$

$\alpha$, $\beta$ are the Courant-Snyder parameters of the magnetized beam, and $A_\pm$ are the 2×2 block matrices of the 4×4 transfer matrix of the RFBT:

$$M_{RFBT} = \frac{1}{2}\begin{bmatrix} A_+ & A_- \\ A_- & A_+ \end{bmatrix} \quad (8)$$

It is apparent that the left side of equation (6) is purely determined by the Courant-Snyder parameters of the initial CAM-dominated beam, while the right side of equation (6) is only determined by the RFBT settings. Since the energy spread of the beam is dominated by the correlated energy spread, the chromatic effect could be compensated if a proper longitudinal correlation of the Courant-Snyder parameters could be added to the initial beam. The focusing element of the transfer matrix for a pure π-mode standing wave cavity is given as [25]:

$$C_{21} = -\frac{eE_0}{\sqrt{2}\gamma_i m_e c^2}\left[\cos^2(\Delta\varphi) + \frac{1}{2}\right]\sin\mu \quad (9)$$

where

$$\mu = \frac{1}{2\sqrt{2}\cos(\Delta\varphi)}\ln\left(\frac{\gamma_f}{\gamma_i}\right) \quad (10)$$

and $m_e$ is the mass of an electron, $c$ is the velocity of light, $E_0$ is the peak electric field of the cavity, $\Delta\varphi$ is the phase of particle with respect to the maximum acceleration phase, $\gamma_i$ and $\gamma_f$ are the Lorenz factor at the entrance and the exit of the cavity respectively. When the beam is off-crest accelerated in the cavity, the difference of the focusing element for electrons being accelerated at different phases will be greatly enhanced if $\gamma_i$ is small, i.e., the longitudinal correlation of the Courant-Snyder parameters could be stronger when the beam is imparted the LPS chirp at lower energy. Detailed simulation results of the chromaticity compensation are demonstrated in Sec. IV.

### III ASTA INJECTOR SETUP

Our simulations consider the ASTA photoinjector currently in its commissioning phase. This 50-MeV injector will support some advanced accelerator R&D including an experiment on beam-driven acceleration in a dielectric-lined asymmetric waveguide. The photoinjector is diagrammed in Fig. 1. The electron source is a 1-1/2 cell cylindrical-symmetric RF gun with a $Cs_2Te$ photocathode. The cathode is illuminated with a 3-ps ultraviolet (UV, $\lambda$=263 nm) laser pulse with uniform radial distribution and a Gaussian temporal profile. The photocathode drive laser enables the generation of a train of bunches repeated at 3 MHz within a 1-ms-duration macropulse. The ~5-MeV electron bunches exiting the RF gun are then accelerated with two superconducting radio-frequency (SRF) TESLA-type cavities (CAV1 and CAV2) up to 50 MeV. A 3rd harmonic cavity (CAV39) operating at 3.9 GHz will eventually be added to linearize the LPS and thereby enhance the beam compression at the bunch compressor. The linearizing cavity is not included in our current simulation. Downstream of this accelerating section the beamline includes quadrupoles and steering dipole magnets along with a bunch compressor (BC1) arranged as a magnetic chicane. Three of the quadrupole magnets upstream of BC1 are skewed and are part of the RFBT (during the flat-beam generation all the normal quadrupole magnets between the skew-quadrupole magnets are turned off). Therefore in the latter configuration, an incoming angular-momentum-dominated beam can be transformed into a flat beam with very high transverse emittance ratio prior to compression in BC1. The 3.5-m-long BC1 consists of four 0.26-m-long rectangular dipole magnets (D1, D2, D3, D4) with bending angles of (+,-,-,+) 18°. The dipole magnets are tilted by 9°. The longitudinal dispersion of BC1 is $R_{56}\approx0.19$ m.

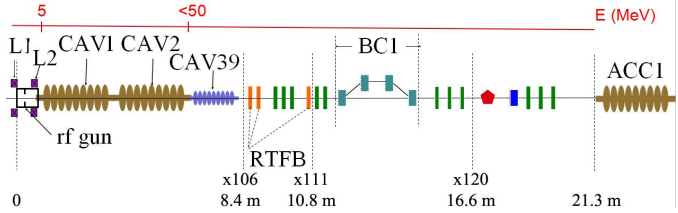

FIG. 1. Overview of the ASTA photoinjector. L1 and L2 correspond to solenoidal lenses, CAV1, 2 and CAV39 are respectively 1.3- and 3.9-GHz SRF accelerating cavities. The blue rectangles represent dipole magnets while the green and orange rectangles are respectively normal and skew quadrupole magnets. BC1 indicates the bunch compressor beamline and ACC1 is an accelerating cryomodule located downstream of the injector for further acceleration to ~300 MeV.

## IV. FLAT BEAM GENERATION

The numerical simulations presented in this Section were carried with ASTRA [26] and IMPACT-T [27]. ASTRA was used to simulate the beam dynamics of the magnetized round beam from its formation at the photocathode up to the entrance of the RFBT (location X106 in Fig. 1) with 10,000 macro particles to obtain the optimized injector settings. We used the cylindrical-symmetric space-charge algorithm included in ASTRA. Further simulations of the RFBT were conducted with IMPACT-T using a 3-D space charge algorithm. The start-to-end simulations were also done with IMPACT-T using 200,000 macro particles and the optimized settings.

### A. Optimization procedure description

A series of simulations were performed with ASTRA to optimize the ASTA injector setups to produce flat bunch with charges of 3.2 nC, 1.0 nC and 20 pC. The photocathode laser is taken to follow a Gaussian distribution with rms duration $\sigma_t = 3$ ps. A uniform transverse laser intensity distribution was taken at the photocathode surface. An initial kinetic energy of 0.55 eV was used to simulate a thermal emittance contribution due to photoemission from the $Cs_2Te$ cathode [28].

The figure-of-merit $\varepsilon_{FOM} = \varepsilon_{sub}^2 / 2\ell$ was minimized at the location of the first skew quadrupole magnet by tuning the launch phase of the RF gun, the laser spot size, the peak fields of the main (L2) and bucking (L1) solenoidal lenses. The quantity $\varepsilon_{sub}$ represent the uncorrelated transverse emittance computed in the Larmor's frame, e.g., after removal of the coupling between the two transverse phase spaces. Thus according to equation (2), $\varepsilon_{FOM}$ should be close to $\varepsilon_{eigen,-}$, as also observed in our simulations.

Once the optimized settings are obtained, the beam is transported through the RFBT. The strengths of the skew quadrupole magnets were determined as follows. First, an analytic solution approximating the quadrupole magnets with their thin-lens matrix was used to estimate the required strengths [29]. This solution was then used as a starting point to search the quadrupole-magnet settings (using the thick-lens matrices) that minimize the figure-of-merit $FOM^2 = \Sigma_{13}^2 + \Sigma_{14}^2 + \Sigma_{23}^2 + \Sigma_{24}^2$, where $\Sigma_{ij}$ is the $(ij)$th element of the 4×4 covariance matrix of the beam at the entrance of the first skew quadrupole. The resulting quadrupole-magnet strengths were then used in IMPACT-T to track the beam evolution throughout the RFBT.

### B. Emittance growth mitigation for RFBT

An LPS chirp of around $h = -5.4$ m$^{-1}$ for full compression is produced off-crest operation of the cavity(ies) upstream of BC1. The optimized settings for the full-compressed flat beams with different final energies were found using the aforementioned procedure. During these optimizations, the peak axial electric field of the gun cavity and CAV2 was fixed at 40 MV/m (eventually up to 45 MV/m in the user phase) and 42 MV/m (the maximum capability of CAV2) respectively to maximize the energy gain, while the peak field of CAV1 and the phase of CAV1 or CAV2 were adjusted to alter the beam energy and match the required LPS chirp. Under nominal operation (e.g. for round-beam generation), only CAV2 is operated off-crest acceleration to provide the necessary correlated fractional energy spread. Off-crest operation of CAV1 is more complex as it can also introduce some compression via velocity bunching. Figure 2 compares the flat-beam normalized vertical emittance and the normalized eigen-emittance for bunches with different charges at the exit of the RFBT with either CAV1 or CAV2 operated off-crest to impart the correlated energy spread required for full compression in BC1. Here, the normalized eigen-emittance $\varepsilon_{n,eigen,-}$ is computed at the entrance of the RFBT, and $\varepsilon_{n,y}$ denotes the normalized vertical emittance of the flat beam at the exit of the RFBT. Taking the 3.2-nC-bunch results as the example, we observed that $\varepsilon_{n,eigen,-}$ remains constant as the beam energy is varied while $\varepsilon_{n,y}$ grows dramatically as the energy increases. This is in fact due to dependence of ponderomotive focusing on peak accelerating field which results in an increase of the transverse beam size at the entrance of the RFBT as the energy increases; see Fig. 3. In the case when the correlated fractional energy spread is imparted by off-crest operation of CAV1, the value of $\varepsilon_{n,\text{eigen},-}$ upstream of the RFBT appears larger than when CAV2 was operated off-crest. The achieved final flat-beam vertical emittance $\varepsilon_{n,y}$ is however much smaller. It is also found that $\varepsilon_{n,y}$ drops below $\varepsilon_{n,\text{eigen},-}$ when the beam energy is lower than ~38.6 MeV. The comparison of the normalized slice vertical emittances of the ~37-MeV, 3.2-nC flat-beams with CAV1 and CAV2 operated off-crest respectively is shown in Fig. 4. The slice emittances computed at both ends of the bunch with CAV2 operated off-crest are much larger than at the center of the bunch because of the chromatic aberration. However, when CAV1 is operated off-crest, it is obvious that the slice vertical-emittance growth at both ends of the bunch is substantially suppressed. The observation depicted in Fig. 2 and 4 confirms our initial suggestion of Sec. II: applying an LPS chirp at lower energy could impart stronger longitudinal correlation of the Courant-Snyder parameters along the bunch. So of these correlations can be removed after the RFBT thereby compensating the chromatics occurring during the flat-beam production process..

Similar results were obtained for bunches with charges of 1.0 nC and 20 pC. Since the chromatic aberration for bunches with lower charges is not obvious, the compensation relative effect decreases as the bunch charge (and associated emittances) becomes smaller. Note that the smallest achievable vertical emittance of the flat beam is the non-correlated eigen-emittance of the CAM-dominated beam, e.g., the eigen-emittance obtained when CAV2 is operated off-crest. It is noteworthy to stress that in our configuration simulations predict a 20-pC bunch to reach 15 nm normalized vertical emittance.

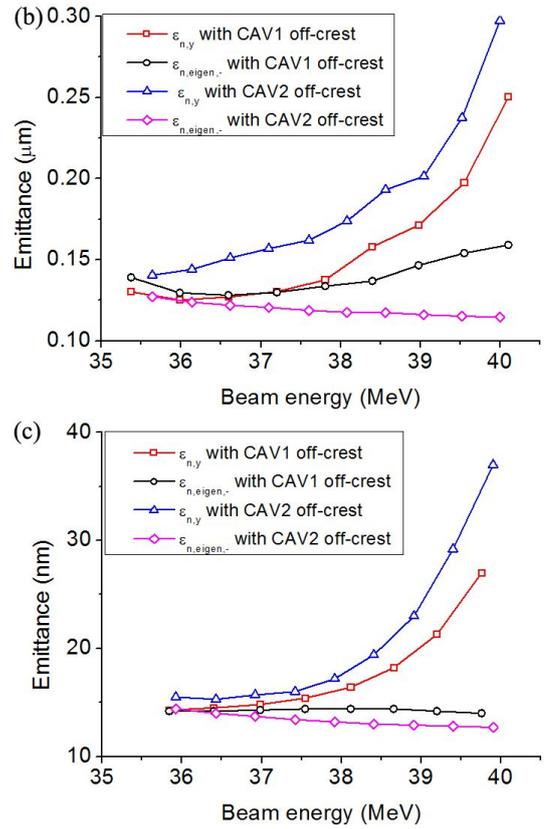

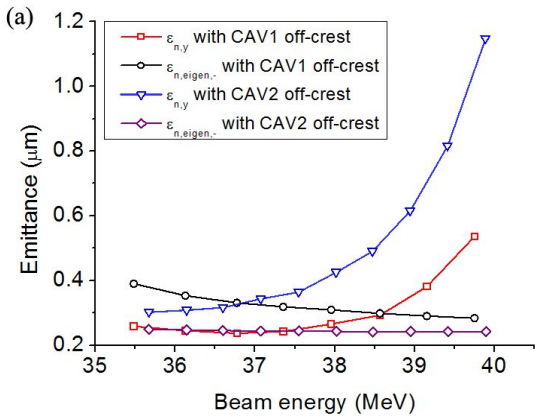

FIG. 2. $\varepsilon_{n,\text{eigen},-}$ (circle) and $\varepsilon_{n,y}$ (rectangular) with CAV1 operated off-crest, and $\varepsilon_{n,\text{eigen},-}$ (diamond) and $\varepsilon_{n,y}$ (triangle) with CAV2 operated off-crest, as a function of beam energy for the beam with bunch charge of (a) 3.2 nC, (b) 1.0 nC, and (c) 20 pC respectively. Note that the unit of the emittance for the 20 pC beam is nm.

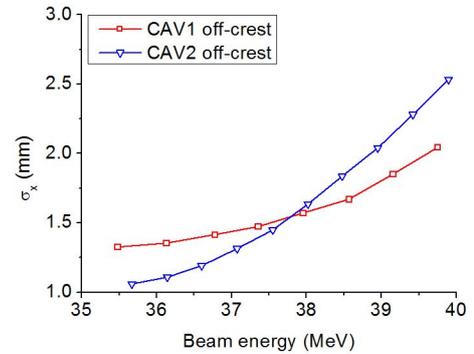

FIG. 3. Transverse beam size at the entrance of the RFBT as a function of the beam energy (corresponding to the results in Fig. 2 (a)).

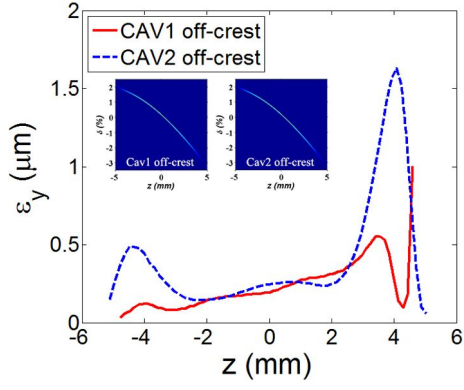

FIG. 4. Comparison of the normalized slice vertical emittances of the ~37 MeV, 3.2-nC flat-beams with CAV1 (red) and CAV2 (blue) operated off-crest respectively. The corresponding longitudinal phase-spaces are also given. The slice plot is produced with 200,000 macro particles in ASTRA.

Another important component of our study was to explore the influence of the RFBT-beamline layout on the smaller flat-beam (vertical) emittance. As an example we consider in Fig. 5 and 6 the case of a 37-MeV 3.2-nC bunch. Here $D_2$ denotes the distance between the first two skew quadrupole magnets and $D_T$ denotes the total length of the RFBT. Normally, the location of the second skew quadrupole magnet is been concerned for a RFBT with a given total length. The simulation results indicate that the first two skew-quadrupole magnets should be as close as possible. The r.m.s. transverse beam sizes evolution along the RFBT beamline are given in Fig. 7 for $D_2/D_T=0.2$ and 0.7 respectively. Since the vertical beam size at the third quadrupole magnet is extremely small, according to equation (5), the third quadrupole magnet will contribute little to the emittance growth caused by the chromatic aberration. When the distance between the first two skew quadrupole magnets increases, the transverse beam sizes at the second quadrupole magnet increase considerably, this results in a larger emittance growth. Note that the integral quadrupole strength of the second skew-quadrupole magnet is weakly dependent on the $D_2/D_T$ ratio. Considering the case when $D_2$ is fixed, the integral quadrupole strengths of all the skew-quadrupole magnets decrease as $D_T$ increases. This decrease in strength mitigates the emittance growth.

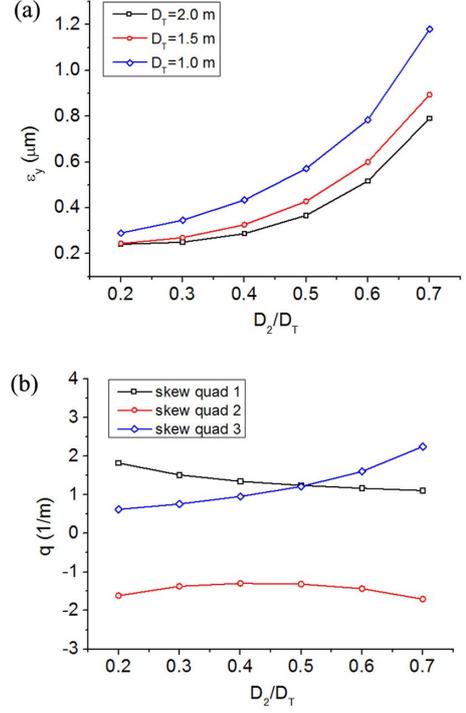

FIG. 5. Vertical emittance of the 37-MeV, 3.2-nC flat beam with different total lengths (a) and the corresponding integral quadrupole strengths for $D_T=2.0$ m (b), as a function of $D_2/D_T$.

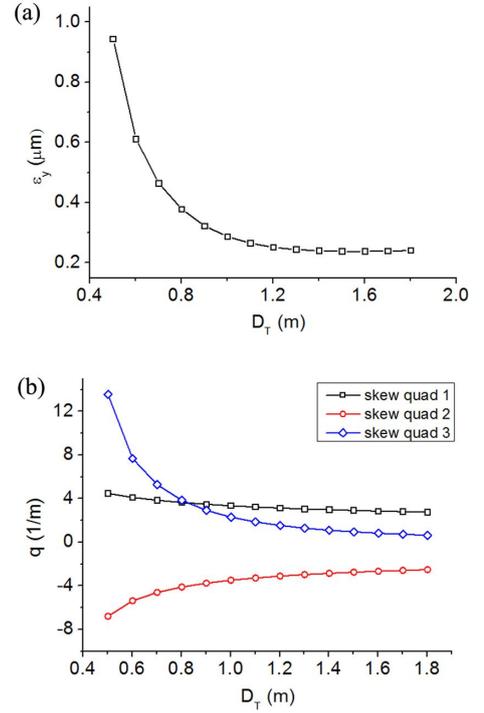

FIG. 6. Vertical emittance of the 37-MeV, 3.2-nC flat beam (a) and the corresponding integral quadrupole strengths (b) as a function of $D_T$ for $D_2=0.2$ m.

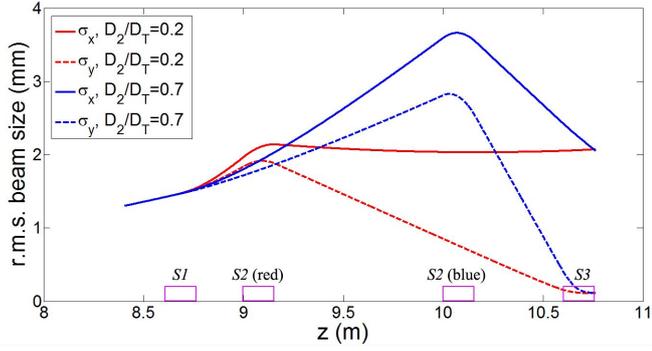

FIG. 7. Transverse beam sizes (37-MeV, 3.2-nC) evolution along the RFBT beamline for $D_2/D_T$=0.2 (red) and 0.7 (blue) with $D_T$=2 m. The S1, S2, S3 magenta rectangles indicate the location of the skew quadrupole magnets. The solid (resp. dash) lines correspond to the horizontal (resp. vertical) r.m.s. beam size values.

### C. Summary of the optimized flat beams

The optimized simulation results for various bunch charges are summarized in Table I. Evolutions of the corresponding transverse emittances from the photocathode to downstream the RFBT transformer for beams with different charges are shown in Fig. 8. These results were simulated with IMPACT-T using the injector-parameter settings optimized with ASTRA. The simulations were in overall agreement and only a -0.5° phase shift for CAV1 in IMPACT-T was required to be added to provide LPS chirp. The phase spaces for the optimized flat beams are given in Fig. 9. ASTRA and IMPACT-T results on the achieved beam parameters all agree to within 5%. In addition, by turning off the space charge effect in IMPACT-T, we observed that the vertical emittance growth of the flat beam due to the space charge effect during RFBT is within 5%, and the horizontal emittance growth of the flat beam due to the space charge effect is negligible (within 0.1%).

Table I. Optimized properties of the flat beams and corresponding injector settings

| Bunch charge (nC) | 3.2 | 1.0 | 0.02 |
|---|---|---|---|
| Energy (MeV) | 36.7 | 36.0 | 36.9 |
| $\varepsilon_{n,x}$ (μm) | 95.4 | 46.9 | 4.32 |
| $\varepsilon_{n,y}$ (μm) | 0.25 | 0.13 | 0.015 |
| Emittance ratio | 382 | 350 | 288 |
| Laser spot size (mm) | 1.26 | 0.93 | 0.25 |
| Gun phase (degree) | 0.0 | 2.0 | 0.0 |
| Peak field of main solenoid (T) | 0.1468 | 0.1486 | 0.1400 |
| Peak field of buck solenoid (T) | 0.1120 | 0.0972 | 0.1220 |
| Peak field of CAV1 (MV/m) | 27.0 | 25.0 | 24.0 |
| Phase of CAV1 (degree) | -48.1 | -47.3 | -36.9 |

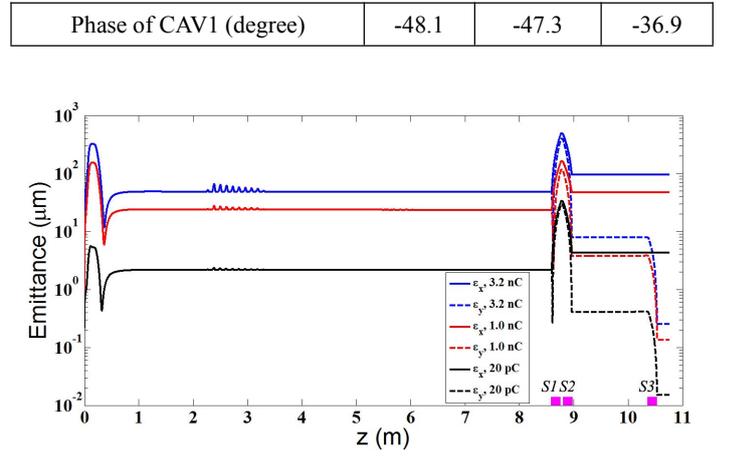

FIG. 8. Evolutions of the normalized transverse projected emittances from the photocathode to X111 for 3.2 (blue), 1.0 (red) and 0.02 nC (black) bunch charge. The S1, S2, S3 magenta rectangles indicate the location of the skew quadrupole magnets. The solid (resp. dash) lines correspond to the horizontal (resp. vertical) emittance values.

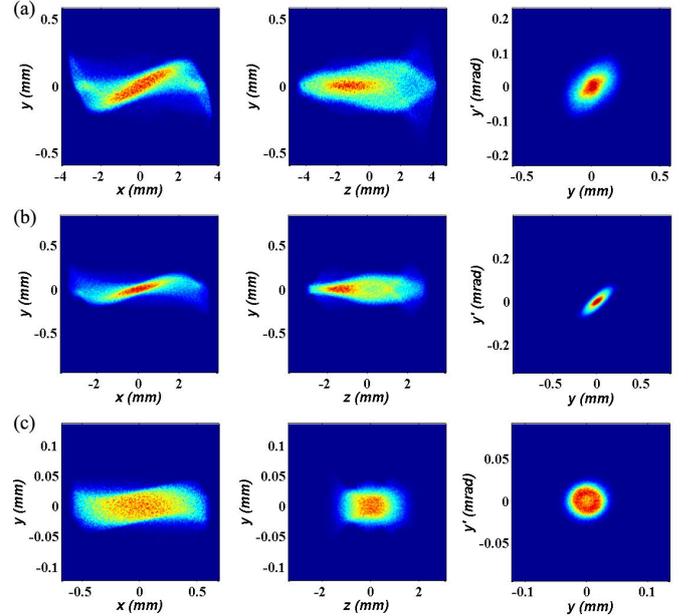

FIG. 9. Simulated $x$-$y$, $z$-$y$ and $y$-$y'$ phase-space at X111 for beams with bunch charges of (a) 3.2 nC; (b) 1.0 nC and (c) 20 pC, respectively

It is noted that the flat beams in Fig. 9 are tilted, especially in the 3.2-nC case. However, after the 3.2-nC flat beam drifts for a few meters, the bunch seems to become flatter, as shown in Fig. 10. This observation indicates that the optimized flat beam cannot be simply judged by its transverse distribution in experiments.

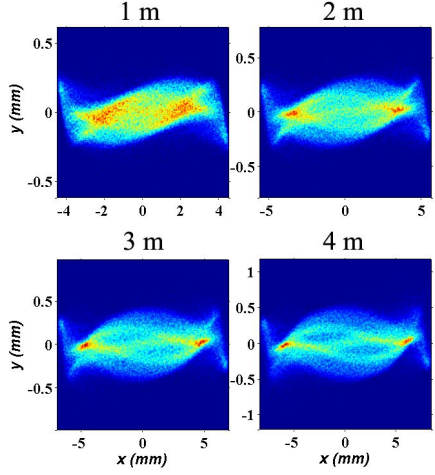

FIG. 10. Evolution of the transverse ($x,y$) bunch distribution of the 3.2-nC flat beam along a drift downstream of the RFBT section. The distances correspond to the drifting distance referenced from X111. The initial distribution at X111 appears in Fig. 9 (a).

## V. FLAT BEAM COMPRESSION

The dynamics of the flat beam in BC1 was simulated with the IMPACT-T program. IMPACT-T notably include fast high-order algorithm to accurately compute the 1D longitudinal CSR force [30]. The layout of BC1 is diagrammed in Fig. 11. The magnetic field profile of the dipole magnet used in IMPACT-T was fitted from the measured data, as shown in Fig. 12. Since the CAV39 cavity was not included in our simulations, the temporal distribution associated with the fully-compressed bunch consists of a high charge concentration in the bunch head with a long trailing tail. The asymmetric character of distribution comes from the quadratic distortion imparted on the LPS during acceleration in CAV1 and CAV2 on the longitudinal phase space. Our choice not to include CAV39 was to accurately model a planned experiment to explore the compression of flat beam for which CAV39 will not be available.

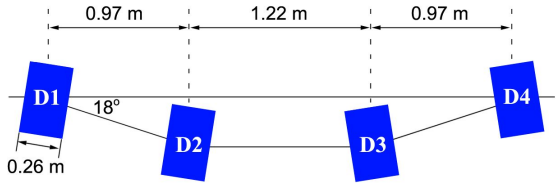

FIG. 11. Layout of BC1

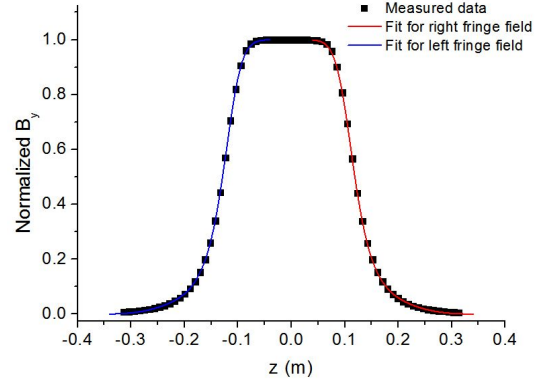

FIG. 12. Measured and fitted field data of the dipole magnet

### A. Bunch length evolution in the chicane

When compressing the round beam in a magnetic chicane, the minimum bunch length should be achieved after the third dipole magnet [9]. However, it is found that the minimum bunch length of the flat beam is always reached after the fourth dipole magnet.

To first order, the r.m.s. bunch length after the third dipole magnet is given as (Appendix B):

$$\sigma_{z_3} = \sqrt{\begin{array}{l}\left(1+hR_{56}^{(3)}\right)^2 \sigma_{z_0}^2 + R_{56}^{(3)2}\sigma_{\delta_u}^2 + \\ \varepsilon_x(R_{52}^{(3)2}\frac{1+\alpha_{x0}^2}{\beta_{x0}} + R_{51}^{(3)2}\beta_{x0} - 2R_{51}^{(3)}R_{52}^{(3)}\alpha_{x0})\end{array}} \quad (11)$$

where $R_{51}^{(3)}$, $R_{52}^{(3)}$, $R_{56}^{(3)}$ are the elements of the transfer matrix from the entrance of the chicane to the exit of the third dipole magnet, $\varepsilon_x$ is the geometric horizontal emittance of the beam, $\beta_{x0}$, $\alpha_{x0}$ are the horizontal Courant-Snyder parameters of the beam at the entrance of the chicane, $\sigma_{\delta u}$ is the r.m.s. uncorrelated fractional energy spread and $\sigma_{z0}$ is the r.m.s. bunch length at the entrance of the chicane.

Finally, the final bunch length after the last dipole magnet for full compression is:

$$\sigma_{z_4} = \sqrt{\left(1+hR_{56}\right)^2 \sigma_{z_0}^2 + R_{56}^2 \sigma_{\delta_u}^2} \quad (12)$$

which gives a full-compression length of $\sigma_{z4}=R_{56}\sigma_{\delta u}$.

It is apparent that the horizontal emittance of the flat beam is much larger than the emittance of the round beam. Therefore, the third term inside the radical sign on the right side of equation (11) will dominate in the bunch length of the flat beam after the third dipole magnet. Considering the case the 37-MeV 3.2-nC flat beam, we have $\sigma_{z0}\approx2.1$ mm, $\sigma_{\delta u}\approx1.8\times10^{-3}$ and $\gamma\varepsilon_x\approx95$ μm. The ratio between the bunch lengths after the third dipole magnet and the fourth dipole magnet as a function of the initial horizontal Courant-Snyder parameters of the flat beam is shown in Fig. 13. It is found that a large initial betatron function and convergent angle help to get better compression of the bunch after the third dipole magnet, but the bunch length is still much longer than the full-compressed condition. However, such optics will result in large vertical beta function inside the chicane, which will cause enormous vertical emittance growth due to chromatic aberration. For example, the vertical emittance of a 3.2-nC flat beam with initial Courant-Snyder parameters $\beta_{x,0}=12$ and $\alpha_{x,0}=3$ increases by a factor of 3 without taking into account any collective effect after passing through the chicane. This observation is in clear contrast with the round-beam case.

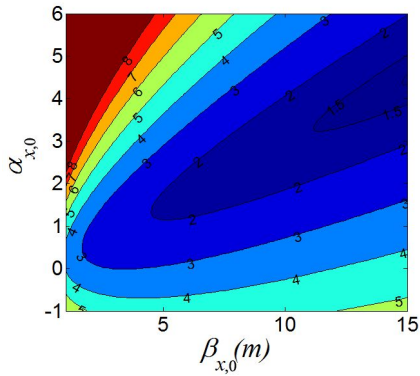

FIG. 13. Ratio between the bunch length after the third dipole magnet and the fourth dipole magnet ($\sigma_{z3}/\sigma_{z4}$) as a function of the initial horizontal Courant-Snyder parameters at the entrance of the chicane. The normalized horizontal emittance is taken to be $\gamma\varepsilon_x \approx 95$ μm.

Two typical evolutions of the bunch length for the 3.2-nC flat beam are shown in Fig. 14. It is found that the bunch can be either compressed or de-compressed after the third dipole magnet with proper initial Courant-Snyder parameters at the entrance of the chicane.

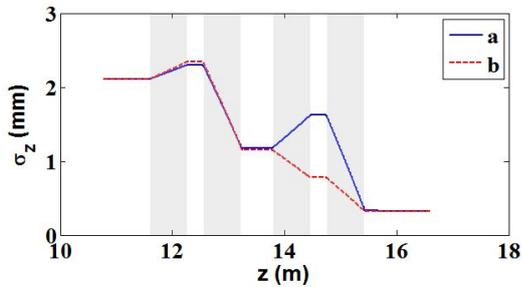

FIG. 14. Bunch length evolutions along BC1 for two typical sets of Courant-Snyder parameters [case a: ($\beta_{x,0}=4.0$, $\alpha_{x,0}=-1.1$) and case 'b': ($\beta_{x,0}=7.0$, $\alpha_{x,0}=1.0$)]. The bunch charge is 3.2 nC and the initial LPS chirp is $h\approx-5.4$m$^{-1}$. The normalized horizontal emittance is taken to be $\gamma\varepsilon_x \approx 95$ μm. The shaded areas indicate the locations of the dipole magnets

### B. Emittance growth in the flat beam compression

For the round beam compression in a magnetic chicane, the majority of the publications only focused on the emittance growth in the bending plane. While in the experiments at CTF-II [32] and LCLS-I [33], both the horizontal and the vertical emittances were measured downstream of the chicane. It was found that the horizontal emittance growth was much larger than the vertical emittance growth when the final bunch length was the shortest. In the following, we will always refer to the projected emittance, unless specified otherwise. With a quadrupole-magnet doublet matching the flat beam into the chicane, a scan of the quadrupole strengths was performed to seek the optimized optical settings for the flat beam compression. The simulation results are summarized in Fig. 15 for a 3.2-nC bunch charge. Fig. 15 reports the horizontal and the vertical betatron functions at the exit of the fourth dipole magnet, the horizontal and vertical relative emittance growth, as well as the peak current, of the compressed flat beam as a function of the initial horizontal Courant-Snyder parameters at the entrance of the chicane are given. For these two dimensional scan, the initial Courant-Snyder parameters before the doublet were ($\beta_{x,i}$, $\alpha_{x,i}$)=(2.6 m, -0.40) and ($\beta_{y,i}$, $\alpha_{y,i}$)=(2.8 m, -0.50).

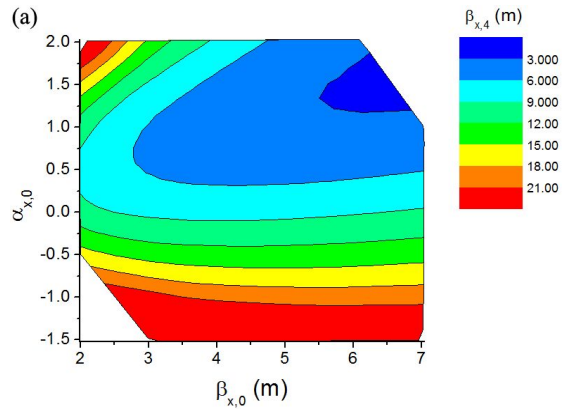

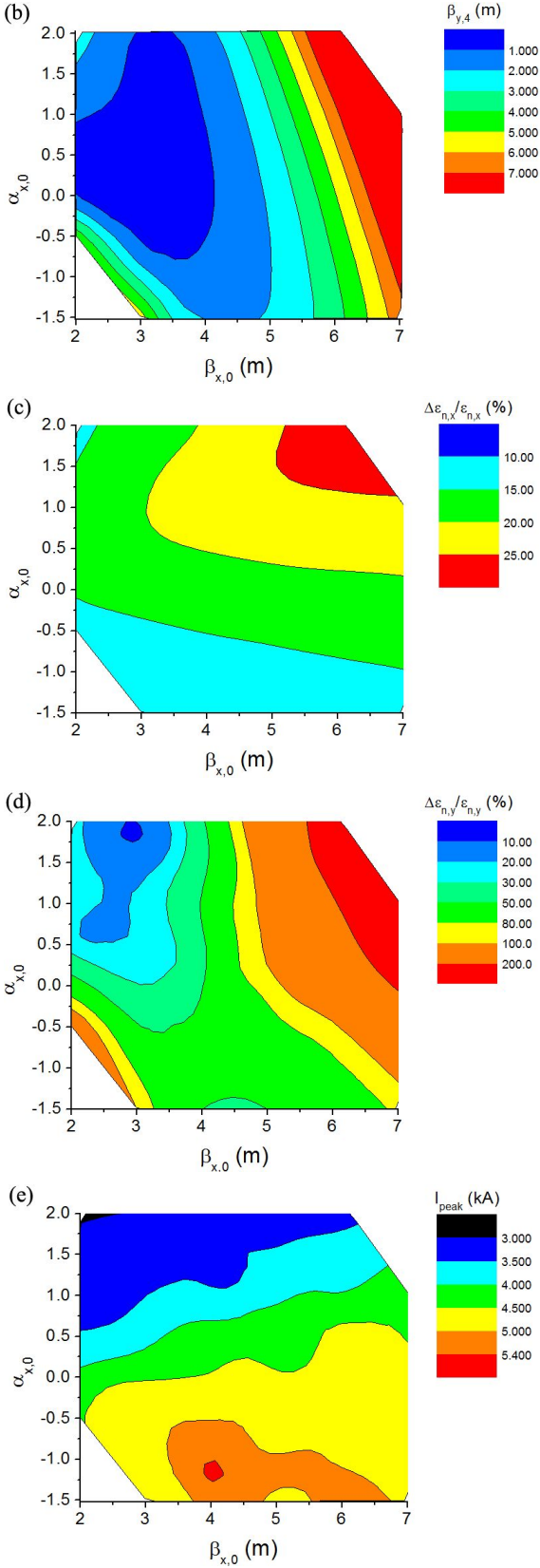

FIG. 15. (a) Horizontal betatron function ($\beta_{x,4}$) at the exit of the fourth dipole magnet; (b) vertical betatron function ($\beta_{y,4}$) at the exit of the fourth dipole magnet; (c) horizontal emittance growth at X120; (d) vertical emittance growth at X120; (e) peak current at X120 as a function of horizontal Courant-Snyder parameters ($\beta_{x,0}$ and $\alpha_{x,0}$) 0.25 m upstream of the chicane.

It is found that the relative horizontal emittance dilution is generally between 10% and 25% for the range of parameter considered. It is well-known that the shift of the slice centroids is the main reason behind the growth of the projected emittance in a bend [34]. The emittance growth over the dipole magnet can be simply estimated by considering the angular spread induced by the energy spread due to CSR effect, which is also demonstrated to be dominated in the emittance growth [35]:

$$\varepsilon_{x,f} \approx \sqrt{\varepsilon_{x,i}^2 + \varepsilon_{x,i}\beta_{x,f}\left\langle \Delta x^{'2} \right\rangle} \quad (13)$$

where $\varepsilon_{x,i}$ and $\varepsilon_{x,f}$ are the initial and final geometric horizontal emittances, $\beta_{x,f}$ is the horizontal betatron function at the bend exit. In the case of the Gaussian current profile, the rms energy loss per unite length per particle due to steady state CSR is given as [36]:

$$\Delta E_{rms}(s) \approx 0.0197 \frac{Ne^2}{\varepsilon_0 \rho^{3/2} \sigma_z(s)^{4/3}} \quad (14)$$

where $s$ denotes the longitudinal position on the design orbit, $N$ is the number of particles in the bunch, $\varepsilon_0$ is the permittivity of free space. Thus the energy spread induced rms angular spread over the dipole magnet can be estimated to be:

$$\Delta x_{rms}^{'} = \frac{1}{E} \int_0^{\rho\theta} R_{26}(s \to \rho\theta) \Delta E_{rms}(s) ds \quad (15)$$

where $E$ is the total energy of the beam and $R_{26}$ is the transfer matrix element associated to the dipole magnet. In the case that $\sigma_z(s)$ is constant and making the small angler and linear optic assumption, equation (15) reduces to

$$\Delta x_{rms}^{'} \approx \frac{\theta \Delta E_{rms}}{2E} \quad (16)$$

which differs from Dohlus [36] by the factor of 2. This apparent discrepancy comes from the assumption that an "instantaneous" CSR-induced energy spread at the entrance of the dipole magnet is assumed by Dohlus. We instead assume the bunch-length evolution inside the fourth dipole magnet is a linear function of the longitudinal beamline coordinate. The emittance growth resulting from equations (13)-(15) as a function of initial bunch length at the entrance of the fourth dipole magnet appears in Fig. 16. Note that in practice the initial bunch length at the entrance of the fourth dipole magnet is actually correlated to $\beta_{x,f}$ as both quantities depends on $\beta_{x,0}$ the incoming betatron function upstream of the chicane. By inspection of Fig. 16 and 15 it is generally observed that the analytic estimate of the emittance dilution are lower than the simulated one. The reason is that a spike will appear in the current profile at the end of the fourth dipole magnet, and the CSR power will be much higher than a Gaussian bunch with the same rms bunch length [37]. However, the angular spread of the beam will not increase much because $R_{26}$ already becomes small at the end of the fourth dipole magnet.

The above result implies that maximizing the bunch length at the entrance of the fourth dipole magnet could mitigate the CSR-induced energy spread. However, maximizing the bunch length at the entrance of the fourth dipole magnet will result in a large horizontal betatron function at the exit of the fourth dipole magnet, while equation (13) suggests that a small horizontal betatron function after the fourth dipole magnet minimizes the horizontal emittance growth. In fact, the simulation results show that the horizontal emittance growth decreases as the horizontal betatron function at the fourth dipole magnet increases, which means that the bunch length at the entrance of the fourth dipole magnet dominates the emittance growth introduced by CSR-induced energy spread in the flat beam case.

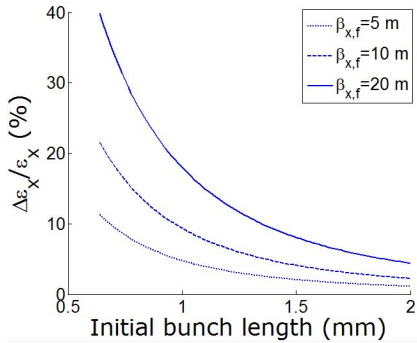

FIG. 16. Horizontal emittance growth as a function of the bunch length at the entrance of the fourth dipole magnet predicted by equation (13)-(15) with $\gamma\varepsilon_{x,i}$=95 μm and $\gamma$=73. The bunch length at the exit of the dipole magnet is 0.32 mm.

The reason that the flat beam could mitigate the emittance growth in the bending plane is summarized as follows. First, the initial horizontal emittance of the flat beam is very large, which dilutes the contribution of the angular spread introduced by the CSR-induced energy spread. Second, the bunch length at the entrance of the fourth dipole magnet is long and the minimum bunch length is only achieved at the end of the fourth dipole magnet, so the angular spread of the beam introduced by the CSR-induced energy spread is not obvious during bending.

Different from the horizontal emittance growth, the vertical emittance only increases after the fourth dipole magnet, as shown in Fig. 17. It implies that the vertical emittance growth is mainly induced by the space charge effect. The CSR effect can also introduce nonlinearity into the particle distribution and increases the emittance growth downstream the chicane. Similar to equation (13), the vertical emittance growth is caused by the space charge induced vertical angular spread:

$$\varepsilon_{y,f}=\sqrt{\varepsilon_{y,i}^2 + \varepsilon_{y,i}\beta_{y,f}\langle\Delta y'^2\rangle} \qquad (17)$$

where $\varepsilon_{y,i}$ and $\varepsilon_{y,f}$ are the initial and final geometric vertical emittances, $\beta_{y,f}$ is the final vertical betatron function. Equation (17) suggests that keeping the vertical betatron function small is crucial to minimize the vertical emittance growth, which is consistent with the simulation results. Therefore, the vertical emittance also benefits from a large horizontal betatron function at the exit of the dipole magnet, which will result in a small vertical betatron function there.

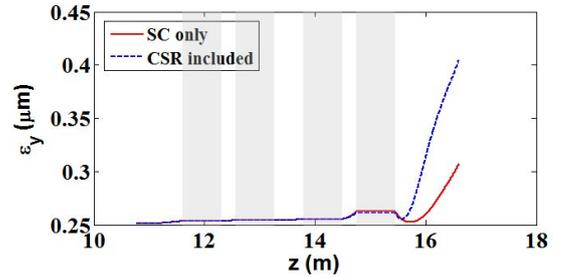

FIG. 17. Vertical emittance evolution of the 3.2-nC flat beam along the BC1 beamline with (blue) and without (red) considering the CSR effect. Initial Courant-Snyder parameters $\beta_{x,0}$=4.0 m, $\alpha_{x,0}$=-1.1 are used. The shaded area indicates the location of the dipole magnets.

The 4-D normalized emittance growth of the 3.2-nC flat beam after compression at the low energy chicane is only found to be about 35%. This is a substantial improvement compared to the round-beam case where the final 4-D emittance was generally found to deteriorate by a factor ~6 when the bunch was fully compressed [17].

For experiments like wakefield-driven acceleration a high peak current is quintessential. Therefore the initial Courant-Snyder parameters settings which produce the highest peak current were selected. The nominal optics is shown in Fig. 18. The peak current of the 3.2-nC bunch is nearly 5.5 kA. Although the final vertical emittance is 60% higher than the initial one, it is only 0.41 μm, which is still extremely low for a full compressed 3.2-nC bunch. It is noted that the horizontal beam size is very large downstream of the chicane but the downstream quadrupole-magnet triplet can be used to focus the beam in both direction to a specific location.

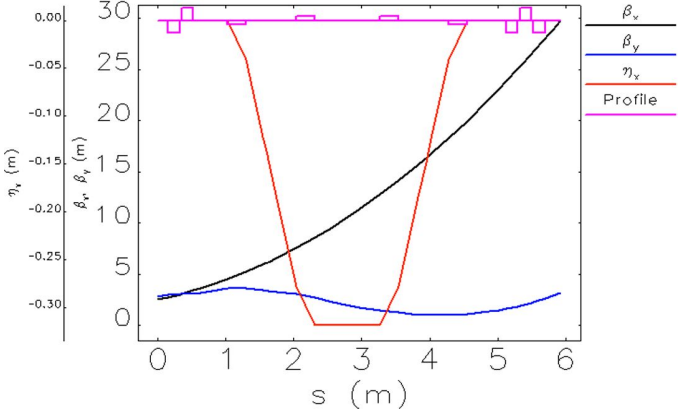

FIG. 18. Nominal optics of the chicane with $\beta_{x,0}=4.0$ and $\alpha_{x,0}=-1.1$. Small purple rectangles located between 1<s<5 m represent dipole magnets, while the larger rectangles correspond to quadrupole magnets. The beam optics is calculated with the ELEGANT code [38]. The quadrupole triplet downstream the chicane (z>5 m) is turned off.

### C. Phase-space dilution during compression

As discussed previously, a large horizontal betatron function at the exit of the fourth dipole magnet is necessary to preserve both the horizontal and the vertical emittances. According to equation (11), there are two ways to maximize the horizontal betatron function at the exit of the fourth dipole magnet. The first solution is to have a horizontally-divergent beam at the entrance of BC1, while the alternative setting corresponds to a convergent beam with a small initial betatron function. However, simulation results show that the peak current of the compressed beam with the second optical solution is much lower than with the former one. The phase-spaces and the current profiles of the beam, with and without including the CSR effect, for the two case of optics mentioned above are displayed in Fig. 19 and Fig. 20 respectively. Some CSR-induced $x$-$z$ phase-space distortions are obvious downstream of BC1. The phase-space distortions with the second optics solution is more serious which results in a large drop of the peak current. Moreover, the $x$-$y$ phase-space distortion may induce asymmetric effect, which is undesired for most experiments.

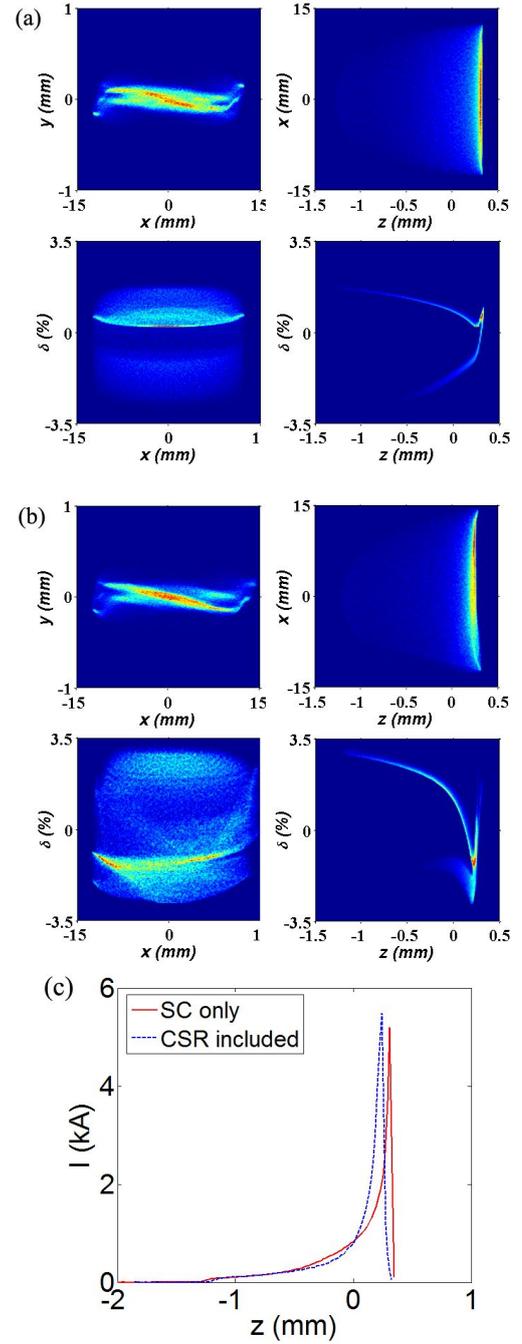

FIG. 19. (a) Beam phase-spaces with only space charge effect; (b) beam phase-spaces with CSR effect; (c) current profiles, at X120 with initial Courant-Snyder parameters $\beta_{x,0}=4.0$ and $\alpha_{x,0}=-1.1$.

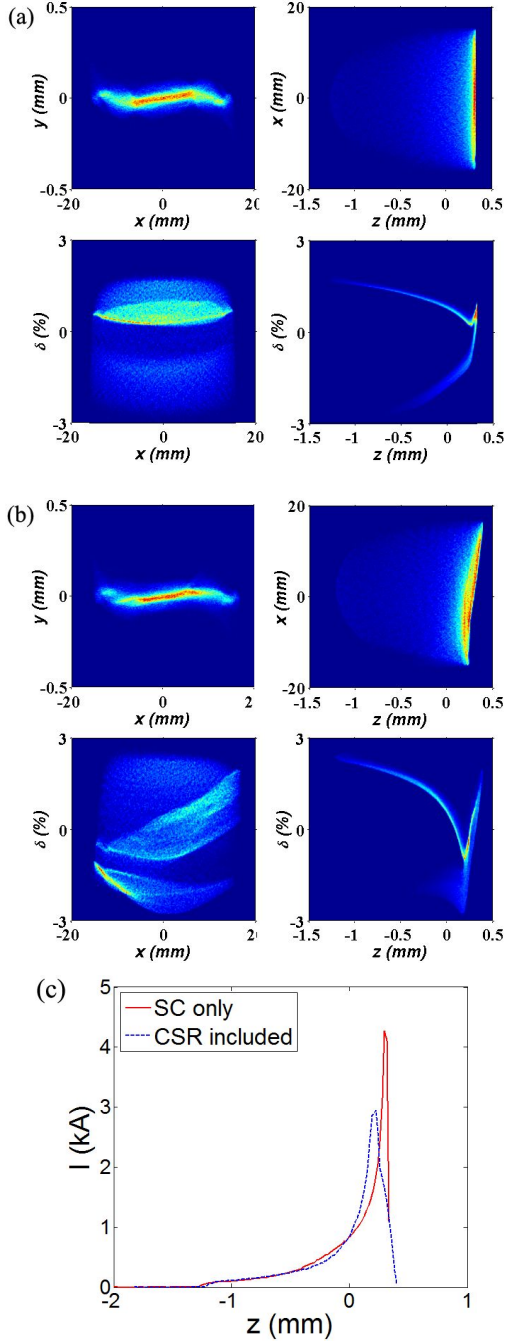

FIG. 20. (a) Beam phase-spaces with only space charge effect; (b) beam phase-spaces with CSR effect; (c) current profiles, at X120 with Courant-Snyder parameters $\beta_{x,0}$=2.0 and $\alpha_{x,0}$=2.0.

A simple possible explanation for the observed distortion in the x-z plane can be elaborated. The longitudinal position of an electron downstream the fourth dipole magnet $z_{4,f}$ is related to its position upstream the fourth dipole magnet $z_{4,i}$ by

$$z_{4,f} = z_{4,i} + \int_0^{\rho\theta} R_{56}(s \to \rho\theta)\delta(s)ds, \quad (18)$$

where

$$R_{56}(s_0 \to s) = \int_{s_0}^{s} \frac{\eta(s')}{\rho(s')} ds'. \quad (19)$$

Since the dispersion function changes with energy, the x-y phase space distortion could be caused if electrons whose final longitudinal positions are supposed to be the same experience different energy loss.

Considering the electrons whose final longitudinal positions are the same, according to equation (B8), the longitudinal position of an electron after the third dipole magnet with respect to the electron at the central orbit is

$$\Delta z_3 = R_{52}^{(3)} x_0' + R_{51}^{(3)} x_0, \quad (20)$$

assuming the bunch length is dominated by the emittance term after the third dipole magnet for the flat beam case, which means that the longitudinal position of these electrons are distributed along the bunch length. Since the energy losses of electrons in a Gaussian-shaped bunch correlate with their longitudinal positions, the dispersion function experienced by these electrons is different during the bending and so do their final longitudinal positions. It is found that there is a strong $x$-$\delta$ correlation for the compressed flat beam with the second optics solution.

### D. Compression of the flat beam with lower charges

The phase-spaces for the optimized 1.0-nC and 20-pC flat beams under full compression are shown in Fig. 21. The properties of the full-compressed flat beams with different charges are summarized in Table II. For the 1.0-nC flat beam, the peak current reaches nearly 2 kA and the x-z phase-space distortion is not obvious. The vertical emittance growth is considerable but the absolute value is still extremely low. The vertical emittance evolution for the 1.0-nC compressed flat beam is shown in Fig. 22. It is apparent that the emittance growth is caused mostly by the space charge effect. For the full-compressed 20-pC flat beam, it is noticeable that there is almost no impact on the emittance and the phase-space due to the collective effects.

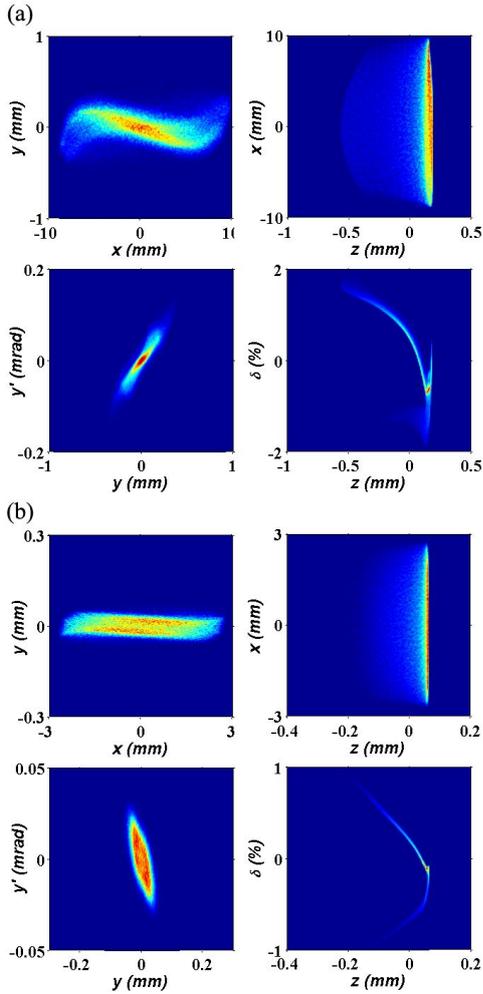

FIG. 21. Phase-spaces of the fully compressed flat beam with bunch charge of (a) 1.0-nC and (b) 20-pC, respectively.

Table II. Properties of the optimized flat beams under full compression

| Bunch charge (nC) | 3.2 | 1.0 | 0.02 |
|---|---|---|---|
| Energy (MeV) | 36.1 | 35.6 | 36.9 |
| $\varepsilon_{n,x}$ (μm) | 107.5 | 50.3 | 4.36 |
| $\varepsilon_{n,y}$ (μm) | 0.41 | 0.20 | 0.016 |
| Peak current (kA) | 5.5 | 1.9 | 0.13 |
| r.m.s. bunch length (mm) | 0.32 | 0.19 | 0.066 |

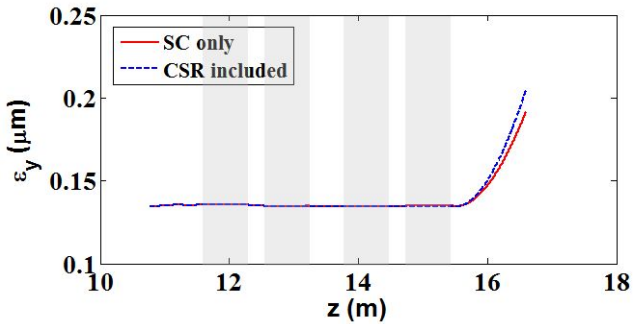

FIG. 22. Vertical emittance evolution of the 1.0-nC flat beam along the beamline with (blue) and without (red) considering the CSR effect. The shaded area indicates the location of the dipole magnets.

## VI. SUMMARY

In this paper, we investigated the formation of compressed flat beams. We especially suggested and verified a scheme to pre-compensate chromatic aberrations occurring in the round-to-flat-beam transformation of beams with large correlated energy spread. We presented start-to-end simulations of flat-beam generation and compression using as an example the 50-MeV photoinjector of the ASTA facility currently in commissioning phase at Fermilab. The projected emittance growth in the bending plane was found to be greatly suppressed with the flat beam due to the fact that the bunch length after the third dipole magnet is much longer than the final bunch length, which reduces the angular spread of the beam introduced by the CSR-induced energy spread in the fourth dipole magnet. For the flat beam, a large horizontal betatron function at the exit of the fourth dipole magnet of the chicane is crucial to preserve both the projected horizontal and the vertical emittances. The high peak current could result in a considerable vertical emittance growth after compression, but the final vertical emittance is still extremely small compared to the round beam case. The optimized normalized vertical emittance of a 3.2-nC bunch with peak current of 5.5 kA is only 0.41 μm. It is noteworthy that the collective effects have almost no influence on the emittance and phase-space of the full-compressed flat beam at low bunch charges. An unprecedented low normalized transverse emittance value of 16 nm was attained in our simulations for a 220-fs and 20-pC compressed electron bunch.

## ACKNOWLEDGEMENTS

The author would like to thank J. Qiang for providing us with a copy of the IMPACT-T code and for fruitful discussions. We are indebted to P. Czerpata, V. Shiltsev, E. Harms and R. Estrada for their support. This work was supported by DOE Contracts No. DE-AC02-07CH11359 with the Fermi Research Alliance, LLC. and No. DE-FG02-08ER41532 with the Northern Illinois University. J. Zhu is supported by a fellowship from the Battelle Memorial Institute.

## APPENDIX A: EMITTANCE GROWTH IN THE RFBT

Under thin lens approximation, the transfer matrix of a skew quadrupole magnet is given as:

$$M_{skew} = \begin{bmatrix} 1 & 0 & 0 & 0 \\ 0 & 1 & q & 0 \\ 0 & 0 & 1 & 0 \\ q & 0 & 0 & 1 \end{bmatrix} \quad (A1)$$

where $q=1/f$ is the integral quadrupole strength, and $f$ is the focal length of the quadrupole magnet. The transverse coordinates of an electron at the exit of a skew quadrupole magnet is related to its coordinates at the entrance of the skew quadrupole magnet by:

$$\begin{bmatrix} x_f \\ x'_f \\ y_f \\ y'_f \end{bmatrix} = \begin{bmatrix} x_i \\ x'_i + qy_i \\ y_i \\ y'_i + qx_i \end{bmatrix} \quad (A2)$$

In this section, the subscripts '$i$' and '$f$' denote the quantities at the entrance and the exit of the skew quadrupole magnet respectively. The vertical geometric emittance change in the $j$th skew quadrupole magnet is then given as:

$$\Delta \varepsilon_{y,j}^2 = \varepsilon_{y,f,j}^2 - \varepsilon_{y,i,j}^2$$
$$= \left( \langle q_j^2 x_{i,j}^2 \rangle \langle y_{i,j}^2 \rangle - \langle q_j x_{i,j} y_{i,j} \rangle^2 \right) \quad (A3)$$
$$+ 2 \left( \langle q_j x_{i,j} y'_{i,j} \rangle \langle y_{i,j}^2 \rangle - \langle q_j x_{i,j} y_{i,j} \rangle \langle y_{i,j} y'_{i,j} \rangle \right)$$

where $\langle \rangle$ defines the second central moment of the particle distribution. The vertical emittance of the final flat beam is:

$$\varepsilon_y^2 = \varepsilon_{eff}^2 + \sum_{j=1}^{3} \Delta \varepsilon_{y,j}^2 \quad (A4)$$

where

$$\varepsilon_{eff}^2 = \varepsilon_u^2 + \ell^2 \quad (A5)$$

The integral quadrupole strength for an electron with momentum $p=p_0(1+\delta)$ is:

$$q = \frac{q_0}{1+\delta} \approx q_0(1-\delta+\delta^2) \quad (A6)$$

where $q_0$ is the integral quadrupole strength of electrons with the momentum $p_0$. Considering there is no correlation between $\delta$ and the transverse coordinates, using the relationship,

$$\langle x_{i,1} y_{i,1} \rangle = 0, \ \langle x_{i,3} y_{i,3} \rangle = \langle x_{f,3} y_{f,3} \rangle = 0 \quad (A7)$$

and substitute equation (A3) and (A6) into (A4), we have

$$\frac{\Delta \varepsilon_y}{\varepsilon_{y0}} \approx \frac{1}{2} \langle \delta^2 \rangle \begin{pmatrix} 2q_{0,1}^2 \langle x_{i,1}^2 \rangle \langle y_{i,1}^2 \rangle + (1+\lambda)q_{0,2}^2 \langle x_{i,2}^2 \rangle \langle y_{i,2}^2 \rangle \\ +2q_{0,3}^2 \langle x_{i,3}^2 \rangle \langle y_{i,3}^2 \rangle - \varepsilon_{eff}^2 \end{pmatrix},$$

(A8)

where $\varepsilon_{y0}$ is the vertical geometric emittance of the flat beam without chromatic aberration, the subscript '$j$' indicates quantities related to the $j$th skew quadrupole magnet and:

$$\lambda = 1 - \frac{\langle x_{i,2} y_{i,2} \rangle^2}{\langle x_{i,2}^2 \rangle \langle y_{i,2}^2 \rangle} \quad (A9)$$

Considering:

$$\langle x_{i,2}^2 \rangle \langle y_{i,2}^2 \rangle \geq \langle x_{i,2} y_{i,2} \rangle^2 \quad (A10)$$

We have:

$$0 \leq \lambda \leq 1 \quad (A11)$$

## APPENDIX B: TRANSFER MATRIX

To the first order, considering the dogleg consists of two symmetric rectangular dipole magnets with bending radius $\rho$ and angle $\theta$ separated by a drift of length $d_1$, the transfer matrices of the dipole magnet and the dogleg are given as [31]:

$$M_{dipole}(\theta, \rho) = \begin{bmatrix} 1 & \rho \sin \theta & 0 & 2\rho \sin^2 \frac{\theta}{2} \\ 0 & 1 & 0 & 2 \tan \frac{\theta}{2} \\ -2 \tan \frac{\theta}{2} & -2\rho \sin^2 \frac{\theta}{2} & 1 & \rho(\sin \theta - \theta) \\ 0 & 0 & 0 & 1 \end{bmatrix}$$

(B1)

$$M_{dog}(L, \eta, \xi) = \begin{bmatrix} 1 & L & 0 & \eta \\ 0 & 1 & 0 & 0 \\ 0 & \eta & 1 & \xi \\ 0 & 0 & 0 & 1 \end{bmatrix} \quad (B2)$$

Respectively, where

$$L = d_1 + 2\rho \sin \theta \quad (B3)$$

$$\eta = 2 \tan \frac{\theta}{2}(d_1 + \rho \sin \theta) \quad (B4)$$

$$\xi = 4 \tan^2 \frac{\theta}{2}(d_1 + \rho \sin \theta) - 2\rho(\theta - \sin \theta) \quad (B5)$$

The transfer matrix of the chicane is then derived as:

$$M_{chicane} = (L, -\eta, \xi) M_{drift}(d_2) M_{dog}(L, \eta, \xi)$$

$$= \begin{bmatrix} 1 & 2L+d_2 & 0 & 0 \\ 0 & 1 & 0 & 0 \\ 0 & 0 & 1 & R_{56} \\ 0 & 0 & 0 & 1 \end{bmatrix} \quad (B6)$$

where $d_2$ is the distance between the two doglegs, and $R_{56}=2\xi$.

For an electron in the beam, its coordinates after the third dipole magnet is related to its coordinates at the entrance of the chicane by:

$$\begin{bmatrix} x_3 \\ x_3' \\ z_3 \\ \delta_3 \end{bmatrix} = M_{dipole}(-\theta,-\rho) M_{drift}(d_2) M_{dog}(L,\eta,\xi) \begin{bmatrix} x_0 \\ x_0' \\ z_0 \\ \delta_0 \end{bmatrix}$$

(B7)

From equation (A7) we have:

$$z_3 = z_0 + R_{56}^{(3)}\delta_0 + R_{52}^{(3)}x_0' + R_{51}^{(3)}x_0 \quad (B8)$$

where

$$R_{51}^{(3)} = 2\tan\frac{\theta}{2} \quad (B9)$$

$$R_{52}^{(3)} = \eta + 2\rho\sin^2\frac{\theta}{2} + 2(L+d_2)\tan\frac{\theta}{2} \quad (B10)$$

$$R_{56}^{(3)} = R_{56} + \rho(\theta - \sin\theta) \quad (B11)$$

$$\delta_0 = \delta_u + hz_0 \quad (B12)$$

where $\delta_u$ is the fractional uncorrelated energy deviation. Further considering there is no correlation between ($x_0$, $x_0'$) and ($z_0$, $\delta_0$), the r.m.s. bunch length after the third dipole magnet is:

$$\sigma_{z_3} = \sqrt{\begin{array}{l}(1+hR_{56}^{(3)})^2\sigma_{z_0}^2 + R_{56}^{(3)2}\sigma_{\delta_u}^2 + \\ \varepsilon_x(R_{52}^{(3)2}\frac{1+\alpha_{x0}^2}{\beta_{x0}} + R_{51}^{(3)2}\beta_{x0} - 2R_{51}^{(3)}R_{52}^{(3)}\alpha_{x0})\end{array}}$$

(B13)

where $\varepsilon_x$ is the geometric horizontal emittance of the beam, $\beta_{x0}$, $\alpha_{x0}$ are the horizontal Courant-Snyder parameters of the beam at the entrance of the chicane, $\sigma_{\delta u}$ is the r.m.s. uncorrelated fractional energy spread and $\sigma_{z0}$ is the r.m.s. bunch length at the entrance of the chicane.